%% file: ms.tex
\documentclass[sigconf]{acmart}

\newcommand{\systemname}{Que2Engage}

\usepackage{algorithm}
\usepackage{multirow}

\AtBeginDocument{%
  \providecommand\BibTeX{{%
    \normalfont B\kern-0.5em{\scshape i\kern-0.25em b}\kern-0.8em\TeX}}}

\copyrightyear{2023}
\acmYear{2023}
\setcopyright{acmlicensed}\acmConference[WWW '23 Companion]{Companion Proceedings of the ACM Web Conference 2023}{April 30-May 4, 2023}{Austin, TX, USA}
\acmBooktitle{Companion Proceedings of the ACM Web Conference 2023 (WWW '23 Companion), April 30-May 4, 2023, Austin, TX, USA}
\acmPrice{15.00}
\acmDOI{10.1145/3543873.3584633}
\acmISBN{978-1-4503-9419-2/23/04}

\begin{document}

\title{{\systemname}: Embedding-based Retrieval for Relevant and Engaging Products at Facebook Marketplace}



\newcommand{\authorStr}{Yunzhong He, Yuxin Tian, Mengjiao Wang, Feier Chen, Licheng Yu, Maolong Tang, Congcong Chen, Ning Zhang, Bin Kuang and Arul Prakash}

\author{Yunzhong He$^{\dagger1}$, Yuxin Tian$^{\dagger2}$, Mengjiao Wang$^1$, Feier Chen$^1$, Licheng Yu$^1$, Maolong Tang$^1$, Congcong Chen$^1$, Ning Zhang$^1$, Bin Kuang$^1$ and Arul Prakash$^1$}
\affiliation{%
  \institution{$^1$Meta, $^2$University of California, Merced}
  \country{}
}

\email{{yunzhong, mengjiaow, feche, lichengyu, mltang, chencc, ningzhang, binkuang, arulprakash}@meta.com}
\email{ytian8@ucmerced.edu}

\renewcommand{\shortauthors}{Yunzhong He}

\begin{abstract}
\input{abstract}
\end{abstract}

\ccsdesc[500]{Information systems~Retrieval models and ranking}
\ccsdesc[500]{Applied computing~Online shopping}

\keywords{Product search, e-commerce, information retrieval}

\maketitle

\section{Introduction}\label{sec:intro}
\input{intro.tex}

\section{Modeling}\label{sec:model}
\input{methology}

\section{Offline Experiments}\label{sec:offline_experiments}
\input{experiments.tex}

\section{Online Experiments}\label{sec:online_testing}
\input{online_testing.tex}

\section{Related Work}\label{sec:related_work}
\input{related_work.tex}

\section{Conclusion}\label{sec:conclusion}
\input{conclusion.tex}

\clearpage
\bibliographystyle{ACM-Reference-Format}
\bibliography{bibliography}

\end{document}

%% file: abstract.tex
Embedding-based Retrieval (EBR) is a powerful search retrieval technique in e-commerce to address semantic matches between search queries and products. However, commerce search engines like Facebook Marketplace Search are complex multi-stage systems with each stage optimized for different business objectives. Search retrieval system usually focuses on query-product semantic relevance, while search ranking puts more emphasis on up-ranking products for high quality engagement. As a result, the end-to-end search experience is a combined result of relevance, engagement, and the interaction between different stages of the system. This presents challenges to EBR systems in optimizing overall search experiences. In this paper we present \systemname, a search EBR system designed to bridge the gap between retrieval and ranking for better end-to-end optimization. \systemname~takes a multimodal \& multitask approach to infuse contextual information into the retrieval stage and balance different business objectives. We show the effectiveness of our approach via a multitask evaluation framework with thorough baseline comparisons and ablation studies. \systemname~ has been deployed into Facebook Marketplace Search engine and shows significant improvements in user engagement in two weeks of A/B testing.

%% file: intro.tex
Embedding-based Retrieval (EBR) has become an important component of e-commerce search engines across Facebook Marketplace, Walmart, Instacart, and more~\cite{ebrtaobao, que2search, walmart, instacart}. In general, EBR models focus on learning embedding representations for search queries and documents, so that documents semantically close to a search query can be retrieved via ANN search \cite{ebrtaobao, que2search, walmart, mobius}. However, search engines are usually complex multi-stage systems optimized for multiple business objectives, so simply optimizing for semantic relevance may not always lead to the best outcome. For example, \cite{que2search} points out that integrating EBR systems can lead to NDCG regressions because downstream re-ranking systems may not always able to rank results retrieved via EBR properly.

In e-commerce platforms like Facebook Marketplace \footnote{www.facebook.com/marketplace}, contextual information such as product price, production condition, seller rating, \emph{etc.} are also important signals to consider to ensure products retrieved are engaging to the searchers. However, we argue that the leverage of contextual information in a search EBR setting towards better searcher engagement is not a trivial problem because (1) traditional EBR modeling techniques based on contrastive learning overly emphasize semantic relevance, so naively apply contextual information in a contrastive learning setting may not work well (2) a product being semantically relevant to a query does not imply that it is engaging to the searcher, and thus simultaneously preserve relevance and engagement is challenging.

In this paper we present \systemname, extending Que2Search~\cite{que2search} to address the aforementioned challenges. It takes a multimodal approach to incorporate contextual signals as a unique modality in its transformer fusion backbone. The model is trained with multitask learning that joins contrastive learning with ranker-style training to not only retrieve semantically relevant products, but also up-ranks the more engaging products like a re-ranking model. Similar to \cite{uniretriever}, we propose a multitask evaluation for EBR models to understand its performances in different domains. We share detailed baseline comparisons and ablation studies using the multitask evaluation framework to illustrate our argument of multi-stage consistency and the effectiveness of our approach in leveraging contextual information.

\systemname~is integrated in Facebook Marketplace Search and powering millions of search queries per day. It has demonstrated significant improvements in searcher engagement via two weeks of online A/B testing.

\renewcommand{\thefootnote}{\fnsymbol{footnote}}
\footnote[0]{$\dagger$ Both authors contributed equally to this research }

%% file: methology.tex
\begin{figure*}[t]
\includegraphics[width=0.9\linewidth]{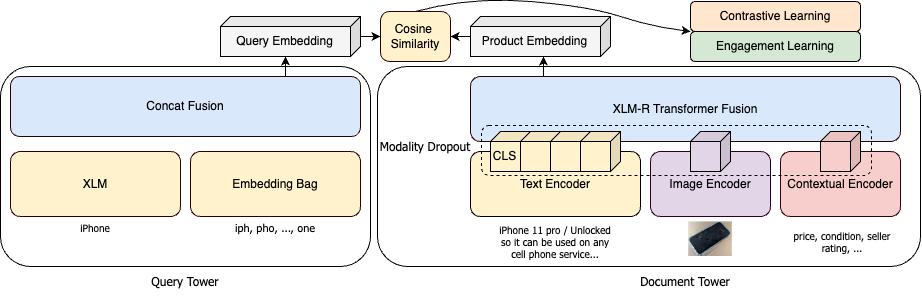}
\centering
\caption{\systemname~architecture overview}
\Description{Architecture overview of Que2Search's two-tower model}
\label{fig:architectures}
\end{figure*}

Figure~\ref{fig:architectures} presents the overall architecture of our \systemname~framework, which is a two-tower neural network consisting of a query and a document tower for learning embedding representations of search queries and e-commerce products, respectively. Multimodal and contextual information of products are fed into the document tower using a transformer-fusion approach, and trained with multi-task learning. In the following, we detail our choices in model architecture and our novel multimodal multitask method.


\subsection{Model Architecture}
\label{sec:arch}

\subsubsection{Query tower}
Similar to \cite{que2search}, we adopt a multi-granular representation of search queries which consists of both raw query text and character trigrams of the query. Raw query text is encoded using a 2-layer XLM~\cite{xlm} encoder, and character trigrams are encoded using an EmbeddingBag \cite{embeddingbag} encoder. Different from \cite{que2search}, we combine the two representations using concatenation instead of attention fusion before sending to the final MLP layer as we find the former yields slightly better performance.


\subsubsection{Contextual information as a modality}
\label{sec:engagement}

For candidate product listings, we use a MLP-based encoder from the document tower to encode their contextual information such as price, category and creation time. Numerical features are represented as single neurons, and categorical features are represented using one-hot encoding. All of the contextual features are concatenated and then fed into a BatchNorm layer followed by a final MLP to ensure a fixed numerical scale and a fixed output length. We call this encoder output a "context token" because it is treated similarly to text and image tokens during the multimodal fusion step covered in section \ref{sec:multimodel_fusion}. Essentially, contextual information is treated as a unique modality in our multimodal framework.

\subsubsection{Multimodal fusion}\label{sec:multimodel_fusion}
Besides encoding the contextual information, we use text encoder to convert the textual fields, \emph{i.e.}, product title and description, into a sequence of word tokens and feed them into the transformer to get the textual embedding. A special [CLS] token is used to encode the whole sentence representation. 
For the variable number of images attached to the document, we take the pre-trained image representations~\cite{groknet} for each of the attached images, apply a shared MLP layer and deep sets~\cite{deep_sets} fusion to get the image dense representation as image modality token. 
We borrow the transformer-fusion architecture used in \cite{commercemm}, where we feed the concatenation of the text tokens, image token as well as context token to the multimodal fusion encoder. Our text encoder and multimodal fusion encoder are initialized from 6-layer XLM-R~\cite{xlmr}, an multilingual language model. As in \cite{commercemm}, the text encoder inherits its first K layers and the multimodal fusion model inherits its remaining M layers. We extract the hidden output of the [CLS] token at the last layer of multimodal fusion encoder and project it to the desired dimension as the final document embedding. 

\subsubsection{Modality dropout}
To ensure that our model does not overly rely on one modality and is robust against missing information during inference time, we introduce a modality dropout mechanism. Specifically, we randomly mask out the output of contextual encoder, image encoder and text encoder with probability of $\delta_{c}$, $\delta_{i}$ and $\delta_{t}$ respectively, and replace the masked ones with tensor of zero.

\subsubsection{Learning image representations}
We explore two variants of image encoders to capture visual information from product images. In method one, we directly apply pre-computed image embedding from the GrokNet model \cite{groknet}, with an MLP layer on top to ensure the image embedding size is consistent all the other tokens in transformer fusion. In method two, we include an off-the-shelf RestNet50 encoder from the CommerceMM model \cite{commercemm} into the document tower, and train the entire document tower as a continued CommerceMM fine-tuning process. While the latter is obviously more powerful because the original image encoder is retained and fine-tuned, the former is much simpler to train and costs less memory in both training and serving. We will compare them in more details in section \ref{sec:offline_experiments}.

\subsection{Multitask Training}
\label{sec:multitask}

\subsubsection{Contrastive learning}\label{sec:contrastive_learning}
We adopt constrastive learning based on batch negative sampling as part of our training objectives, where positive samples are user engaged <query, product> pairs sampled from de-identified and aggregated search log, and negative samples are generated by randomly combining queries and products within a mini-batch of positive samples. Formally, we introduce the relevance loss $L_{relevance}$ as follows 

\begin{equation}
L_{relevance}=\frac{1}{B} \sum_{i=1}^{B}-\log \left\{\frac{\exp \left\{s \cdot \kappa\left(q_{i}, d_{i}\right)\right\}}{\sum_{j=1}^{B} \exp \left\{s \cdot \kappa\left(q_{i}, d_{j}\right)\right\}}\right\}
\label{loss:bnce}
\end{equation}

\noindent where $B$ is the batch size, $\kappa$ is a similarity kernel that is implemented as the cosine similarity, and $s$ denotes a scaling factor which is simply a fixed value $s=20$ throughout all experiments. The value is the same as the Que2Search \cite{que2search} model.

\subsubsection{Learning contextual information}
Although batch negative sampling has proven useful in learning semantic relevance in the search EBR problem \cite{walmart, que2search}, we notice that it is not sufficient in learning contextual information towards user engagement. For example, contextual information like product price is a distinguishing factor to identify engaging products among relevant products (\emph{i.e.} a product can be relevant but receives no engagement because the listed price is not reasonable). 
However, during batch negative training, the model receives little negative supervision from products with very unreasonable prices, since all negative samples are generated from engaged <query, product> pairs.
Fundamentally, as \cite{sampledsoftmaxeffectiveness} points out, this is because batch negative methods implicitly sample from the distribution of engaged products, which may not be the true distribution of the inventory. Methods like mixing random negatives \cite{mixedbatch} and in-batch hard negative mining \cite{que2search} are proposed to mitigate the problem. However, we observe that negative samples generated by those approaches are still too easy for the model to pick up the nuances in contextual information. Therefore, we propose an auxiliary training task that optimizes the model directly towards finding engaging products among relevant product. Specifically, we augment the training set in section \ref{sec:contrastive_learning} by including <query, product> pairs displayed to the searchers but receive no searcher engagements as hard negatives, and computes a BCE loss on those samples. Formally, we define the loss $L_{engagement}$ as follows
\begin{equation}
    L_{engagement} = -{(y_i\log(c_i) + (1 - y_i)\log(1 - c_i))}
\end{equation}
where $c_i = s \cdot \kappa\left(q_{i}, d_{i}\right)$. 

To combine $L_{relevance}$ and $L_{engagement}$, we define the final multitask loss as
\begin{equation}
    L_{}(\theta) = \lambda_1 \cdot L_{relevance} + \lambda_2 \cdot L_{engagement}
\end{equation}
where $\theta$ is the model parameters, $\lambda_1$ and $\lambda_2$ are the weighting parameters chosen empirically.

%% file: experiments.tex
\subsection{Dataset}
We collect 150 million <query, product> pairs displayed to searchers from Facebook Marketplace's search log, with 75 million receiving downstream engagements as positive samples, and the rest of 75 million as negative samples. The data is de-identified and aggregated to remove any personal information before evaluation proceeds. For offline evaluation, we collect 26k human-rated data as the relevance evaluation set. The human-rated dataset is generated by letting raters to decide whether a result is relevant to a query or not. The candidates to be rated are generated by a stratified sampling of the search queries and products, which includes both easy and hard samples. To evaluate user engagement, we reserve one future date among the 150 million de-identified and aggregated search log data.

\subsection{Baselines and Ablation Studies}
We choose Que2Search \cite{que2search} as our baseline model, which is a two-tower model based on attention fusion of pre-trained XLM encoders \cite{xlm} and image representations. We further augment the model with encoders based on contextual information, as well as with mixed batch method \cite{mixedbatch} to incorporate hard negatives from products displayed to searchers. For the treatment group, we use the \systemname~with pre-computed image embeddings, because in practice its simplicity is preferred during the actual model productionalization, and we share the comparison against an alternative image encoder based on fine-tuning of the CommerceMM encoder separately.

\subsection{Experimental Setup}
\subsubsection{Evaluation metrics}
Similar to \cite{uniretriever}, we adopt a multitask evaluation framework to measure semantic relevance and searcher engagement separately. Semantic relevance is measured using the 26k human-rated dataset, and searcher engagement is measured using the 220k future engagement dataset. For both datasets, we rank them using the cosine similarity from our model and report ROC\_AUC as the evaluation metric. Note that one significant difference between the two datasets lies in personalization. For the human-rated dataset, raters are asked to make judgement based purely on an objective guideline around textual and visual relevance (\emph{e.g.} whether there is a catalog or brand mismatch between a query and a product), while more subjective factors outside of the guideline (\emph{e.g.} whether the listed price or product condition is appealing) play important roles in the engagement dataset. 

Therefore, ROC\_AUC on the relevance dataset measures how well the model predicts search relevance (similar to the in relevance degree in \cite{uniretriever}), which is the main evaluation metric used in \cite{que2search}. We have also found it correlates well with search retrieval performance. ROC\_AUC of the engagement dataset essentially measures its performance on the search ranking task, because the candidates are all products with user impressions. In fact, it is also a good indicator of the consistency between search retrieval and search ranking.

\subsubsection{Parameters}
We develop all of the models on Nvidia A100 GPUs using the PyTorch Multimodal framework \cite{mmf}. Models are trained using batch size of 512, and optimized using Adam optimizer with a learning rate of $4e-4$. We set weighting parameter $\lambda_1$ and $\lambda_2$ as 0.8 and 0.2 respectively. For the modality dropout, $\delta_c$, $\delta_i$ and $\delta_t$ are 0.5, 0 and 0.5. 
We directly feed the text tokens and tokens from other modalities into the multimodal transformer as ~\cite{commercemm}, \emph{i.e.}, our multimodal transformer is an early-fusion model with 0-layer text encoder and 6-layer multimodal fusion encoder. One exception to the aforementioned settings is that when comparing pre-computed image embeddings with fine-tuning the CommerceMM encoder, due to the increased GPU memory consumption of the fine-tuning approach, we adjust the batch size to 64 for that particular experiment. Learning rate was also adjusted to $5e-5$ for the CommerceMM training to avoid NaN in the loss function computation.

\subsection{Results}

\subsubsection{Analysis of baseline results}
The top part of table \ref{res:core} outlines the performance of baseline models. We can see that Que2Search does well on the relevance evaluation but performs poorly on the engagement dataset, which is expected because it is optimized for semantic relevance. While adding contextual information to Que2Search itself does not improve the performance on engagement evaluation, changing training objective to include mixed negatives significantly improves the prediction. Mixed negatives also helps Que2Search to better leverage contextual features towards engagement prediction with row 4 in table \ref{res:core} achieving the highest ROC\_AUC on engagement evaluation. This aligns with our hypothesis in section \ref{sec:contrastive_learning} - vanilla batch negative is insufficient to learn the nuances in engagement prediction from contextual information due to the sampling bias introduced by generating negatives from the positives, and this can be mitigated by introducing mixed negatives. Note that row 4 and 5 in table \ref{res:core} suggests a regression in relevance evaluation along with the improvement of engagement evaluation. This is expected because contextual information like product price and condition are irrelevant to the query-product relevance guideline provided to our raters, and thus we do not expect the leverage of contextual signals to improve the relevance evaluation.

\subsubsection{\systemname~ and ablation studies}
The second part of table \ref{res:core} shows that engagement evaluation can be significantly improved with the \systemname~approach, with the full \systemname~ using multitask learning and modality dropout achieving the best results across the two evaluation methods. Ablation study on the loss function suggests that multitask training leads to the biggest improvement in engagement evaluation, suggesting that a more focused loss function on hard negatives works better than simply mixing the negatives. Finally, modality dropout further improves both metrics and specially the relevance evaluation, suggesting that forcing missing modalities may prevent the model over-fitting on one task and thus regressing the other.

\subsubsection{Image encoders}
We also compare the two approaches to incorporate image signals into the multimodal fusion framework. We can see that although both methods are based on pre-training tasks, being able to fine-tune the image encoder indeed outperforms the original frozen GrokNet \cite{groknet} approach used in Que2Search in both evaluations. However, given that fine-tuning the image encoder end-to-end requires significantly more GPU memory and training time, we do not include this technique in the production model for simplicity. And the ablation study is done with a reduced batch size of 64 which significantly regresses the absolute relevance evaluation because larger batch size is proven helpful in contrastive learning \cite{crossbatch}. However, we hope to productionalize this technique with memory efficiency optimizations such as \cite{crossbatch, uniretriever} and with more powerful hardware.

\begin{table}[t]
\resizebox{1.0\linewidth}{!}
{
\begin{tabular}{lcc}
\toprule
{Model} & Engagement & Relevance  \\ \hline
Que2Search\cite{que2search}    & 55.88      & 67.14                \\
Que2Search w/ contextual encoder   &    55.85             &  66.74               \\
Que2Search w/ mixed batch loss   &    63.63             &  63.79               \\
Que2Search w/ contextual encoder + mixed batch loss & 64.45  & 61.17   \\ \hline
\systemname~w/ mixed batch loss & 64.70 & 60.36 \\
\systemname~w/ multitask training  & 76.13 & 65.63 \\
\systemname ~w/ multitask training + modality dropout   &   \textbf{76.90}  &\textbf{67.21}          \\
\bottomrule
\end{tabular}
}
\caption{\small {Results for baseline comparison and ablation studies}}
\label{res:core}
\vspace{-2.0em}
\end{table}

\begin{table}[t]
\resizebox{0.9\linewidth}{!}
{
\begin{tabular}{lcc}
\toprule
Method  & Engagement & Relevance  \\ \hline
pre-computed image embedding & 74.35 & 60.19 \\ 
fine-tuned CommerceMM encoder & \textbf{74.67} & \textbf{60.55} \\
\bottomrule
\end{tabular}
}
\caption{\small Results for image encoder comparison}
\label{res:image}
\vspace{-2.0em}
\end{table}

%% file: online_testing.tex
We deploy \systemname~on Facebook Marketplace Search as a parallel retrieval source to the traditional lexical-based search retrieval. For online A/B testing, we compare \systemname~with Que2Search \cite{que2search}, which is our previous production model solely optimized for semantic relevance. We measure both NDCG and searcher engagement for this A/B testing. Note that NDCG is calculated from human-rated labels using simulated search results similar to how the human-rated evaluation set is generated. Two weeks of online A/B testing shows that \systemname~improves online searcher engagement by 4.5\% while keeping NDCG neutral, which aligns with our offline multitask evaluation results.

%% file: related_work.tex
In recent years, Embedding-based Retrieval (EBR) has been adopted in e-commerce search to retrieve semantically relevant products as a complement of lexical retrieval \cite{que2search, walmart, ebrtaobao, uniretriever}. Siamese neutral networks \cite{siamese, Tian_2020} trained with contrastive learning loss \cite{sampledsoftmaxeffectiveness, 03809} are among the popular modeling choices for EBR in both search and recommendation systems \cite{crossbatch, mixedbatch, twinbert}. However, naive contrastive learning using in-batch negatives can suffer from missing interesting negative samples \cite{uniretriever, mixedbatch} and being memory-hungry \cite{crossbatch}. Variants of contrastive learnings are proposed to address them by incorporating smarter negative sampling \cite{mixedbatch, uniretriever, que2search} and optimizing memory usage \cite{crossbatch, uniretriever}. Sometimes, teacher-student learning is also used as an auxiliary task to improve relevance \cite{mobius, uniretriever}. In search retrieval, Pre-trained Language Models (PLM) are widely adopted because the main focus of retrieval is often textual relevance \cite{walmart, twinbert}. Yet, \cite{ebrtaobao, que2search, uniretriever} point out that contextual information beyond text relevance and consistency with re-ranking stage are also important factors to EBR systems' end-to-end performance. Recently, \cite{uniretriever} developed a unified training scheme to balance multiple optimization objectives, yet the role of contextual information in the multi-objective setting is rarely discussed.

%% file: conclusion.tex
We present the need of EBR modeling to balance relevance and engagement in the real-world applications like Facebook Marketplace, and introduce \systemname, our latest search EBR system, to address the challenges. Through baseline comparisons and ablations studies, we show the effectiveness of our innovations in incorporating contextual signals, multimodal techniques, representation learning and multitask learning. We have deployed \systemname~ on Facebook Marketplace Search. Through two weeks of A/B testing, we show that it outperforms our existing state-of-the-art search EBR system \cite{que2search} and significantly improved searcher engagement on product listed at Facebook Marketplace. 

%% file: ms.bbl

\begin{thebibliography}{20}


\ifx \showCODEN    \undefined \def \showCODEN     #1{\unskip}     \fi
\ifx \showDOI      \undefined \def \showDOI       #1{#1}\fi
\ifx \showISBNx    \undefined \def \showISBNx     #1{\unskip}     \fi
\ifx \showISBNxiii \undefined \def \showISBNxiii  #1{\unskip}     \fi
\ifx \showISSN     \undefined \def \showISSN      #1{\unskip}     \fi
\ifx \showLCCN     \undefined \def \showLCCN      #1{\unskip}     \fi
\ifx \shownote     \undefined \def \shownote      #1{#1}          \fi
\ifx \showarticletitle \undefined \def \showarticletitle #1{#1}   \fi
\ifx \showURL      \undefined \def \showURL       {\relax}        \fi
\providecommand\bibfield[2]{#2}
\providecommand\bibinfo[2]{#2}
\providecommand\natexlab[1]{#1}
\providecommand\showeprint[2][]{arXiv:#2}

\bibitem[Bell et~al\mbox{.}(2020)]%
        {groknet}
\bibfield{author}{\bibinfo{person}{Sean Bell}, \bibinfo{person}{Yiqun Liu},
  \bibinfo{person}{Sami Alsheikh}, \bibinfo{person}{Yina Tang},
  \bibinfo{person}{Edward Pizzi}, \bibinfo{person}{M. Henning},
  \bibinfo{person}{Karun Singh}, \bibinfo{person}{Omkar Parkhi}, {and}
  \bibinfo{person}{Fedor Borisyuk}.} \bibinfo{year}{2020}\natexlab{}.
\newblock \showarticletitle{GrokNet: Unified Computer Vision Model Trunk and
  Embeddings For Commerce}. In \bibinfo{booktitle}{\emph{Proceedings of the
  26th ACM SIGKDD International Conference on Knowledge Discovery and Data
  Mining}} (Virtual Event, CA, USA) \emph{(\bibinfo{series}{KDD '20})}.
  \bibinfo{publisher}{Association for Computing Machinery},
  \bibinfo{address}{New York, NY, USA}, \bibinfo{pages}{2608–2616}.
\newblock
\showISBNx{9781450379984}
\urldef\tempurl%
\url{https://doi.org/10.1145/3394486.3403311}
\showDOI{\tempurl}


\bibitem[Bromley et~al\mbox{.}(1993)]%
        {siamese}
\bibfield{author}{\bibinfo{person}{Jane Bromley}, \bibinfo{person}{Isabelle
  Guyon}, \bibinfo{person}{Yann LeCun}, \bibinfo{person}{Eduard S\"{a}ckinger},
  {and} \bibinfo{person}{Roopak Shah}.} \bibinfo{year}{1993}\natexlab{}.
\newblock \showarticletitle{Signature Verification Using a "Siamese" Time Delay
  Neural Network}. In \bibinfo{booktitle}{\emph{Proceedings of the 6th
  International Conference on Neural Information Processing Systems}} (Denver,
  Colorado) \emph{(\bibinfo{series}{NIPS'93})}. \bibinfo{publisher}{Morgan
  Kaufmann Publishers Inc.}, \bibinfo{address}{San Francisco, CA, USA},
  \bibinfo{pages}{737–744}.
\newblock


\bibitem[Conneau et~al\mbox{.}(2020)]%
        {xlmr}
\bibfield{author}{\bibinfo{person}{Alexis Conneau}, \bibinfo{person}{Kartikay
  Khandelwal}, \bibinfo{person}{Naman Goyal}, \bibinfo{person}{Vishrav
  Chaudhary}, \bibinfo{person}{Guillaume Wenzek}, \bibinfo{person}{Francisco
  Guzm{\'a}n}, \bibinfo{person}{Edouard Grave}, \bibinfo{person}{Myle Ott},
  \bibinfo{person}{Luke Zettlemoyer}, {and} \bibinfo{person}{Veselin
  Stoyanov}.} \bibinfo{year}{2020}\natexlab{}.
\newblock \showarticletitle{Unsupervised Cross-lingual Representation Learning
  at Scale}. In \bibinfo{booktitle}{\emph{Proceedings of the 58th Annual
  Meeting of the Association for Computational Linguistics}}.
  \bibinfo{publisher}{Association for Computational Linguistics},
  \bibinfo{address}{Online}, \bibinfo{pages}{8440--8451}.
\newblock
\urldef\tempurl%
\url{https://doi.org/10.18653/v1/2020.acl-main.747}
\showDOI{\tempurl}


\bibitem[Conneau and Lample(2019)]%
        {xlm}
\bibfield{author}{\bibinfo{person}{Alexis Conneau} {and}
  \bibinfo{person}{Guillaume Lample}.} \bibinfo{year}{2019}\natexlab{}.
\newblock \bibinfo{booktitle}{\emph{Cross-Lingual Language Model Pretraining}}.
\newblock \bibinfo{publisher}{Curran Associates Inc.}, \bibinfo{address}{Red
  Hook, NY, USA}.
\newblock


\bibitem[Fan et~al\mbox{.}(2019)]%
        {mobius}
\bibfield{author}{\bibinfo{person}{Miao Fan}, \bibinfo{person}{Jiacheng Guo},
  \bibinfo{person}{Shuai Zhu}, \bibinfo{person}{Shuo Miao},
  \bibinfo{person}{Mingming Sun}, {and} \bibinfo{person}{Ping Li}.}
  \bibinfo{year}{2019}\natexlab{}.
\newblock \showarticletitle{MOBIUS: Towards the Next Generation of Query-Ad
  Matching in Baidu's Sponsored Search}. In
  \bibinfo{booktitle}{\emph{Proceedings of the 25th ACM SIGKDD International
  Conference on Knowledge Discovery and Data Mining}} (Anchorage, AK, USA)
  \emph{(\bibinfo{series}{KDD '19})}. \bibinfo{publisher}{Association for
  Computing Machinery}, \bibinfo{address}{New York, NY, USA},
  \bibinfo{pages}{2509–2517}.
\newblock
\showISBNx{9781450362016}
\urldef\tempurl%
\url{https://doi.org/10.1145/3292500.3330651}
\showDOI{\tempurl}


\bibitem[Li et~al\mbox{.}(2021)]%
        {ebrtaobao}
\bibfield{author}{\bibinfo{person}{Sen Li}, \bibinfo{person}{Fuyu Lv},
  \bibinfo{person}{Taiwei Jin}, \bibinfo{person}{Guli Lin},
  \bibinfo{person}{Keping Yang}, \bibinfo{person}{Xiaoyi Zeng},
  \bibinfo{person}{Xiao-Ming Wu}, {and} \bibinfo{person}{Qianli Ma}.}
  \bibinfo{year}{2021}\natexlab{}.
\newblock \showarticletitle{Embedding-Based Product Retrieval in Taobao
  Search}. In \bibinfo{booktitle}{\emph{Proceedings of the 27th ACM SIGKDD
  Conference on Knowledge Discovery and Data Mining}} (Virtual Event,
  Singapore) \emph{(\bibinfo{series}{KDD '21})}.
  \bibinfo{publisher}{Association for Computing Machinery},
  \bibinfo{address}{New York, NY, USA}, \bibinfo{pages}{3181–3189}.
\newblock
\showISBNx{9781450383325}
\urldef\tempurl%
\url{https://doi.org/10.1145/3447548.3467101}
\showDOI{\tempurl}


\bibitem[Liu et~al\mbox{.}(2021)]%
        {que2search}
\bibfield{author}{\bibinfo{person}{Yiqun Liu}, \bibinfo{person}{Kaushik
  Rangadurai}, \bibinfo{person}{Yunzhong He}, \bibinfo{person}{Siddarth
  Malreddy}, \bibinfo{person}{Xunlong Gui}, \bibinfo{person}{Xiaoyi Liu}, {and}
  \bibinfo{person}{Fedor Borisyuk}.} \bibinfo{year}{2021}\natexlab{}.
\newblock \showarticletitle{Que2Search: Fast and Accurate Query and Document
  Understanding for Search at Facebook}. In
  \bibinfo{booktitle}{\emph{Proceedings of the 27th ACM SIGKDD Conference on
  Knowledge Discovery and Data Mining}} (Virtual Event, Singapore)
  \emph{(\bibinfo{series}{KDD '21})}. \bibinfo{publisher}{Association for
  Computing Machinery}, \bibinfo{address}{New York, NY, USA},
  \bibinfo{pages}{3376–3384}.
\newblock
\showISBNx{9781450383325}
\urldef\tempurl%
\url{https://doi.org/10.1145/3447548.3467127}
\showDOI{\tempurl}


\bibitem[Lu et~al\mbox{.}(2020)]%
        {twinbert}
\bibfield{author}{\bibinfo{person}{Wenhao Lu}, \bibinfo{person}{Jian Jiao},
  {and} \bibinfo{person}{Ruofei Zhang}.} \bibinfo{year}{2020}\natexlab{}.
\newblock \showarticletitle{TwinBERT: Distilling Knowledge to Twin-Structured
  Compressed BERT Models for Large-Scale Retrieval}. In
  \bibinfo{booktitle}{\emph{Proceedings of the 29th ACM International
  Conference on Information and Knowledge Management}} (Virtual Event, Ireland)
  \emph{(\bibinfo{series}{CIKM '20})}. \bibinfo{publisher}{Association for
  Computing Machinery}, \bibinfo{address}{New York, NY, USA},
  \bibinfo{pages}{2645–2652}.
\newblock
\showISBNx{9781450368599}
\urldef\tempurl%
\url{https://doi.org/10.1145/3340531.3412747}
\showDOI{\tempurl}


\bibitem[Magnani et~al\mbox{.}(2022)]%
        {walmart}
\bibfield{author}{\bibinfo{person}{Alessandro Magnani}, \bibinfo{person}{Feng
  Liu}, \bibinfo{person}{Suthee Chaidaroon}, \bibinfo{person}{Sachin Yadav},
  \bibinfo{person}{Praveen Reddy~Suram}, \bibinfo{person}{Ajit
  Puthenputhussery}, \bibinfo{person}{Sijie Chen}, \bibinfo{person}{Min Xie},
  \bibinfo{person}{Anirudh Kashi}, \bibinfo{person}{Tony Lee}, {and}
  \bibinfo{person}{Ciya Liao}.} \bibinfo{year}{2022}\natexlab{}.
\newblock \showarticletitle{Semantic Retrieval at Walmart}. In
  \bibinfo{booktitle}{\emph{Proceedings of the 28th ACM SIGKDD Conference on
  Knowledge Discovery and Data Mining}} (Washington DC, USA)
  \emph{(\bibinfo{series}{KDD '22})}. \bibinfo{publisher}{Association for
  Computing Machinery}, \bibinfo{address}{New York, NY, USA},
  \bibinfo{pages}{3495–3503}.
\newblock
\showISBNx{9781450393850}
\urldef\tempurl%
\url{https://doi.org/10.1145/3534678.3539164}
\showDOI{\tempurl}


\bibitem[Paszke et~al\mbox{.}(2017)]%
        {embeddingbag}
\bibfield{author}{\bibinfo{person}{Adam Paszke}, \bibinfo{person}{Sam Gross},
  \bibinfo{person}{Soumith Chintala}, \bibinfo{person}{Gregory Chanan},
  \bibinfo{person}{Edward Yang}, \bibinfo{person}{Zachary DeVito},
  \bibinfo{person}{Zeming Lin}, \bibinfo{person}{Alban Desmaison},
  \bibinfo{person}{Luca Antiga}, {and} \bibinfo{person}{Adam Lerer}.}
  \bibinfo{year}{2017}\natexlab{}.
\newblock \showarticletitle{Automatic differentiation in pytorch}.
\newblock  (\bibinfo{year}{2017}).
\newblock


\bibitem[Singh et~al\mbox{.}(2020)]%
        {mmf}
\bibfield{author}{\bibinfo{person}{Amanpreet Singh}, \bibinfo{person}{Vedanuj
  Goswami}, \bibinfo{person}{Vivek Natarajan}, \bibinfo{person}{Yu Jiang},
  \bibinfo{person}{Xinlei Chen}, \bibinfo{person}{Meet Shah},
  \bibinfo{person}{Marcus Rohrbach}, \bibinfo{person}{Dhruv Batra}, {and}
  \bibinfo{person}{Devi Parikh}.} \bibinfo{year}{2020}\natexlab{}.
\newblock \bibinfo{title}{MMF: A multimodal framework for vision and language
  research}.
\newblock
  \bibinfo{howpublished}{\url{https://github.com/facebookresearch/mmf}}.
\newblock


\bibitem[Tian et~al\mbox{.}(2020)]%
        {Tian_2020}
\bibfield{author}{\bibinfo{person}{Yuxin Tian}, \bibinfo{person}{Xueqing Deng},
  \bibinfo{person}{Yi Zhu}, {and} \bibinfo{person}{Shawn Newsam}.}
  \bibinfo{year}{2020}\natexlab{}.
\newblock \showarticletitle{Cross-Time and Orientation-Invariant Overhead Image
  Geolocalization Using Deep Local Features}. In
  \bibinfo{booktitle}{\emph{Proceedings of the IEEE/CVF Winter Conference on
  Applications of Computer Vision (WACV)}}.
\newblock


\bibitem[Tian et~al\mbox{.}(2022)]%
        {03809}
\bibfield{author}{\bibinfo{person}{Yuxin Tian}, \bibinfo{person}{Shawn Newsam},
  {and} \bibinfo{person}{Kofi Boakye}.} \bibinfo{year}{2022}\natexlab{}.
\newblock \bibinfo{title}{Image Search with Text Feedback by Additive Attention
  Compositional Learning}.
\newblock
\newblock


\bibitem[Wang et~al\mbox{.}(2021)]%
        {crossbatch}
\bibfield{author}{\bibinfo{person}{Jinpeng Wang}, \bibinfo{person}{Jieming
  Zhu}, {and} \bibinfo{person}{Xiuqiang He}.} \bibinfo{year}{2021}\natexlab{}.
\newblock \showarticletitle{Cross-Batch Negative Sampling for Training
  Two-Tower Recommenders}. In \bibinfo{booktitle}{\emph{Proceedings of the 44th
  International ACM SIGIR Conference on Research and Development in Information
  Retrieval}} (Virtual Event, Canada) \emph{(\bibinfo{series}{SIGIR '21})}.
  \bibinfo{publisher}{Association for Computing Machinery},
  \bibinfo{address}{New York, NY, USA}, \bibinfo{pages}{1632–1636}.
\newblock
\showISBNx{9781450380379}
\urldef\tempurl%
\url{https://doi.org/10.1145/3404835.3463032}
\showDOI{\tempurl}


\bibitem[Wu et~al\mbox{.}(2022)]%
        {sampledsoftmaxeffectiveness}
\bibfield{author}{\bibinfo{person}{Jiancan Wu}, \bibinfo{person}{Xiang Wang},
  \bibinfo{person}{Xingyu Gao}, \bibinfo{person}{Jiawei Chen},
  \bibinfo{person}{Hongcheng Fu}, \bibinfo{person}{Tianyu Qiu}, {and}
  \bibinfo{person}{Xiangnan He}.} \bibinfo{year}{2022}\natexlab{}.
\newblock \bibinfo{title}{On the Effectiveness of Sampled Softmax Loss for Item
  Recommendation}.
\newblock
\newblock
\showeprint[arxiv]{2201.02327}~[cs.IR]


\bibitem[Xie et~al\mbox{.}(2022)]%
        {instacart}
\bibfield{author}{\bibinfo{person}{Yuqing Xie}, \bibinfo{person}{Taesik Na},
  \bibinfo{person}{Xiao Xiao}, \bibinfo{person}{Saurav Manchanda},
  \bibinfo{person}{Young Rao}, \bibinfo{person}{Zhihong Xu},
  \bibinfo{person}{Guanghua Shu}, \bibinfo{person}{Esther Vasiete},
  \bibinfo{person}{Tejaswi Tenneti}, {and} \bibinfo{person}{Haixun Wang}.}
  \bibinfo{year}{2022}\natexlab{}.
\newblock \showarticletitle{An Embedding-Based Grocery Search Model at
  Instacart}.
\newblock \bibinfo{journal}{\emph{arXiv preprint arXiv:2209.05555}}
  (\bibinfo{year}{2022}).
\newblock


\bibitem[Yang et~al\mbox{.}(2020)]%
        {mixedbatch}
\bibfield{author}{\bibinfo{person}{Ji Yang}, \bibinfo{person}{Xinyang Yi},
  \bibinfo{person}{Derek~Zhiyuan Cheng}, \bibinfo{person}{Lichan Hong},
  \bibinfo{person}{Yang Li}, \bibinfo{person}{Simon Wang},
  \bibinfo{person}{Taibai Xu}, {and} \bibinfo{person}{Ed~H. Chi}.}
  \bibinfo{year}{2020}\natexlab{}.
\newblock \showarticletitle{Mixed Negative Sampling for Learning Two-tower
  Neural Networks in Recommendations}.
\newblock


\bibitem[Yu et~al\mbox{.}(2022)]%
        {commercemm}
\bibfield{author}{\bibinfo{person}{Licheng Yu}, \bibinfo{person}{Jun Chen},
  \bibinfo{person}{Animesh Sinha}, \bibinfo{person}{Mengjiao Wang},
  \bibinfo{person}{Yu Chen}, \bibinfo{person}{Tamara~L. Berg}, {and}
  \bibinfo{person}{Ning Zhang}.} \bibinfo{year}{2022}\natexlab{}.
\newblock \showarticletitle{CommerceMM: Large-Scale Commerce MultiModal
  Representation Learning with Omni Retrieval}. In
  \bibinfo{booktitle}{\emph{Proceedings of the 28th ACM SIGKDD Conference on
  Knowledge Discovery and Data Mining}} (Washington DC, USA)
  \emph{(\bibinfo{series}{KDD '22})}. \bibinfo{publisher}{Association for
  Computing Machinery}, \bibinfo{address}{New York, NY, USA},
  \bibinfo{pages}{4433–4442}.
\newblock
\showISBNx{9781450393850}
\urldef\tempurl%
\url{https://doi.org/10.1145/3534678.3539151}
\showDOI{\tempurl}


\bibitem[Zaheer et~al\mbox{.}(2017)]%
        {deep_sets}
\bibfield{author}{\bibinfo{person}{Manzil Zaheer}, \bibinfo{person}{Satwik
  Kottur}, \bibinfo{person}{Siamak Ravanbakhsh}, \bibinfo{person}{Barnabas
  Poczos}, \bibinfo{person}{Russ~R Salakhutdinov}, {and}
  \bibinfo{person}{Alexander~J Smola}.} \bibinfo{year}{2017}\natexlab{}.
\newblock \showarticletitle{Deep Sets}. In \bibinfo{booktitle}{\emph{Advances
  in Neural Information Processing Systems}},
  \bibfield{editor}{\bibinfo{person}{I.~Guyon}, \bibinfo{person}{U.~Von
  Luxburg}, \bibinfo{person}{S.~Bengio}, \bibinfo{person}{H.~Wallach},
  \bibinfo{person}{R.~Fergus}, \bibinfo{person}{S.~Vishwanathan}, {and}
  \bibinfo{person}{R.~Garnett}} (Eds.), Vol.~\bibinfo{volume}{30}.
  \bibinfo{publisher}{Curran Associates, Inc.}
\newblock
\urldef\tempurl%
\url{https://proceedings.neurips.cc/paper/2017/file/f22e4747da1aa27e363d86d40ff442fe-Paper.pdf}
\showURL{%
\tempurl}


\bibitem[Zhang et~al\mbox{.}(2022)]%
        {uniretriever}
\bibfield{author}{\bibinfo{person}{Jianjin Zhang}, \bibinfo{person}{Zheng Liu},
  \bibinfo{person}{Weihao Han}, \bibinfo{person}{Shitao Xiao},
  \bibinfo{person}{Ruicheng Zheng}, \bibinfo{person}{Yingxia Shao},
  \bibinfo{person}{Hao Sun}, \bibinfo{person}{Hanqing Zhu},
  \bibinfo{person}{Premkumar Srinivasan}, \bibinfo{person}{Denvy Deng},
  \bibinfo{person}{Qi Zhang}, {and} \bibinfo{person}{Xing Xie}.}
  \bibinfo{year}{2022}\natexlab{}.
\newblock \bibinfo{title}{Uni-Retriever: Towards Learning The Unified Embedding
  Based Retriever in Bing Sponsored Search}.
\newblock
\newblock
\showeprint[arxiv]{2202.06212}~[cs.IR]


\end{thebibliography}
